\documentclass[12pt]{JHEP3}
\usepackage{epsfig}
\preprint{ {\tt hep-th/0303013} \\ {IC/2003/13} }
\newcommand{\be}[1]{ \begin{equation}\label{#1} }
\newcommand{\ee}{\end{equation}}
\newcommand{\bea}[1]{\begin{eqnarray}\label{#1} }
\newcommand{\eea}{\end{eqnarray}}
\newcommand{\eq}[1]{(\ref{#1})}

\newcommand{\uu}{{\rm{u}}}
\newcommand{\vv}{{\rm{v}}}
\newcommand{\x}{{\rm{x}}}
\newcommand{\y}{{\rm{y}}}
\title{Plane waves with weak singularities} 
\author{Justin R. David \\ High Energy Section, \\
The Abdus Salam International Centre for Theoretical Physics, 
\\Strada Costiera, 11-34014 Trieste, Italy.\\
\email{justin@ictp.trieste.it} }
\abstract{ 
We study a class of time dependent solutions 
of the vacuum Einstein equations which are plane waves 
with weak null singularities. 
This singularity is weak in the sense that 
though the tidal forces diverge at the singularity, 
the rate of divergence is such that the distortion suffered by
a freely falling observer remains finite.
Among such weak singular plane waves there is a sub-class which do not exhibit
large back reaction in the presence of test scalar probes.
String propagation in these backgrounds is smooth
and there is a natural way  
to continue the  metric beyond the
singularity. This continued metric admits string propagation
without the string becoming infinitely excited. 
We construct a one parameter family of smooth metrics which are at a
finite distance in the space of metrics from the extended metric and
a well defined operator in the string sigma model 
which resolves the singularity.}

\begin{document}
\baselineskip 3.5ex
\section{Introduction}

There has been a lot of interest in time dependent backgrounds in string
theory. This has been mainly motivated by the desire to understand 
the cosmological singularity or the singularity behind horizons of  black holes
within string theory. These singularities are space like unlike the 
time like orbifold singularities which have been well understood in string
theory. The simplest time dependent backgrounds which can be constructed in
string theory 
are  orbifolds involving either a boost or a null boost 
which  have space like singularities and null singularities
respectively at their fixed point 
\cite{Horowitz:1991ap,
Balasubramanian:2002ry,Nekrasov:2002kf,
Simon:2002ma,Liu:2002ft}. 
Though these backgrounds are simple to construct and one can quantize the string
modes they have been shown to be  potentially
unstable to formation of space like singularities like that of
the ones behind horizons of black holes when probed by test particles. 
In \cite{Horowitz:2002mw} 
it was shown that the introduction of a single particle in the
covering space of orbifolds by null boosts 
causes the space to collapse to a space like
singularity, in \cite{Lawrence:2002aj} it was shown how 
a  homogeneous energy distribution in the transverse directions causes the
singularity in the orbifolds by null boosts to become space like. 
and in \cite{Liu:2002kb} 
it was argued that this instability was reflected in the
divergence of string scattering amplitudes.  
Other time dependent backgrounds constructed involved
orbifolding with boosts and a translation in an orthogonal direction 
\cite{Cornalba:2002fi}
and  orbifolding with null boosts and a translation
\cite{Fabinger:2002kr,Liu:2002kb}. For the case or orbifolding with a
boosts and a translation the singularity is time like and is resolved
in string theory by orientifold planes \cite{Cornalba:2002nv} and 
they are shown to be stable to perturbations in
\cite{Cornalba:2003ze}.
Other simple time dependent backgrounds were studied in 
\cite{Elitzur:2002rt,Craps:2002ii,Buchel:2002kj, 
Berkooz:2002je}

Perhaps the next simplest backgrounds involving null singularities in string
theory are exact plane waves with null singularities. 
Their metric is given by 
\be{p-wave}
ds^2 = dudv + F_{ij}(u) x^i x^j du^2 + dx^i dx^i.
\ee
If $F_{ij}$ is traceless they are solutions to the vacuum Einstein equations
and for ${\rm Tr} F \neq 0$ they are solutions in string theory if they are
supported by appropriate  antisymmetric tensor-form field strengths  
or a non constant dilaton. These exact plane wave backgrounds are singular
from the general relativity point of view 
if $F_{ij}$ diverges at some point $u=u_0$ 
(see for instance in \cite{Marolf:2002bx,Brecher:2002bw}) 
String propagation on these backgrounds were explored in 
\cite{Gueven:1987ad,Amati:1989sa, Horowitz:1990bv,Horowitz:1990sr, deVega:1992ke}
and more recently, 
\cite{Papadopoulos:2002bg,Blau:2002js} 
have investigated a plane wave in which the
classical string equations of motion can be solved explicitly.

As plane waves with null singularities are not obtained by an orbifold of
Minkowski space they do not posses the instability found in 
\cite{Horowitz:2002mw} for the case of oribolds
involving boosts and null boosts. 
They are exact solutions of string theory to all orders in $\alpha'$
\cite{Amati:1989sa,Horowitz:1990bv}. The presence of the covariantly
constant null Killing vector 
ensures absence of particle production \cite{Gibbons:1975jb} in these
backgrounds and 
that they preserve half the supersymmetry when embedded
in super string theory.
This makes them good candidates to study the issue of null singularities in
string theory though in general the string modes cannot be quantized explicitly.

In this paper we study a class of plane waves which are solutions of the vacuum
Einstein equations with a weak null singularity.
They are weak in the sense defined in \cite{Ellis,Tipler},  the tidal forces
diverge at the singularity but the distortion caused by them on an object freely
falling into the singularity
remains finite.  Among these weak singularities we find a sub-class
in which the back reaction on a test scalar probe remains finite at the
singularity. This and the fact that the singularity is weak ensures that 
classical string modes 
and their first derivative  remains finite at the singularity.
Thus there is a natural way to continue them across the singularity 
by matching the modes and their first derivative 
across the singularity to extend the metric beyond
it.  

In string theory, for a plane wave to be non-singular not only should 
the string modes be extended across the singularity but also the 
the string should not become infinitely excited as it passes
through the singularity \cite{Horowitz:1990bv,Horowitz:1990sr}. 
For instance in the case of 
exact plane waves which  are vacuum solutions of Einstein's
equations with a delta function singularity (eg. $F_{ij} = \epsilon_{ij}
\delta(u)$ ) the string modes can be continued across the singularity
but the string gets infinitely excited as
it passes through it and therefore this plane wave is singular from
the point of view of string theory.
In the case of weak singular plane waves which are studied in this paper we show
that the string modes continued across the singularity are not infinitely
excited, thus justifying the extended metric.
As further support for the extended metric we show that there exists a
family of smooth metrics which are at a finite distance in the space of metrics
from the extended metric. We demonstrate 
the existence of a deformation which smoothes out the singularity.  

The paper is organized as follows. In section 2 we 
introduce the backgrounds we will study in this paper and review the notion of a
weak singularity. We show that among weak singular plane waves there is a subset
for which the energy of a test scalar particle does not diverge at the
singularity ensuring that the back reaction is mild. 
In section 3 we study string propagation in this background,
we show that the classical string modes 
and their first  derivative is finite at the singularity,
this allows one to smoothly continue them across the singularity.
In section 4 we show that the string does not get infinitely excited as it
passes through the singularity in the extended metric. Then we show that there
are smooth metrics which are a finite distance in the moduli space from the
extended metric and the existence of a well defined operator in the string
sigma model which resolves the singularity.
In section 5 we state our conclusions.

\section{Plane waves with  weak singularities}

In this section we review the conditions of when a plane wave is singular and
demonstrate 
when they have weak singularities. We then show that among the  class of weakly
singular plane waves there is a subset in which the stress energy tensor of a
test scalar does not diverge at the singularity.

\subsection{A singular plane wave}

For the purposes of this paper we restrict our attention 
to the following metric
\be{epwave}
ds^2 = dudv + F(u) (x^2 - y^2) du^2 + dx^2 + dy^2,
\ee
these are solutions to the vacuum Einstein equations.
The considerations in this paper can be generalized to other plane wave
solutions.  $F(u)$ in \eq{epwave} 
is any function which vanishes  at infinity and singular at $u=0$.
For example $F(u) = e^{-u}/u^\alpha$ with $\alpha>0$ satisfies this criteria.
The metric is defined for $u>0$, the geodesic equations for 
this metric is given by
\bea{geo}
& &\frac{d^2 u}{d\lambda^2} = 0, \\ \nonumber
& &\frac{d^2 x}{d\lambda^2} -F(u) x (\frac{du}{d\lambda})^2= 0, \\ \nonumber
& &\frac{d^2 y}{d\lambda^2} + F(u) y (\frac{du}{d\lambda})^2 = 0,  \\ \nonumber
& &\frac{d}{d\lambda} \left( \frac{d u}{d\lambda} \frac{dv}{d\lambda} +
(\frac{d x}{d\lambda})^2 + (\frac{d y}{d\lambda})^2 + 
F(u) (x^2-y^2) (\frac{du}{d\lambda})^2  \right) =0,
\eea
where $\lambda$ refers to the affine parameter.
It is clear that from these equations that 
if there is a singularity in $F(u)$ at $u=0$ 
the space time exists only for $u >0$
as every time like geodesic with $u = -p \lambda$ , $\lambda \in (-\infty,
0)$, reaches the singularity at $\lambda =0$.
The point $u=0$ is not a scalar curvature singularity as all curvature
invariants vanish for the  plane wave metric. One can  characterize the
singular behaviour of the curvature tensor in a coordinate invariant way as
follows. Consider a time like geodesic which ends at the singularity, 
an orthonormal frame which is parallel propagated 
along this geodesic is given by
\bea{orth-frame}
e_0 = ( p, \dot{v}, \dot{x}, \dot{y} ), \quad &\quad&\quad
e_1 = ( p, \dot{v} + \frac{2}{p}, \dot{x}, \dot{y} ), \\ \nonumber
e_2 = (0, -\frac{2 \dot{x}}{p}, 1, 0 ) , \quad&\quad&\quad
e_3 = (0, -\frac{2 \dot{y}}{p}, 0, 1).
\eea
Here $p$ is the velocity for the $u$ coordinate, 
the vector $e_0$ is the tangent to the geodesic and normalized as $e_0^\mu
e_{0\mu} =-1$, the dot refers to derivative with respect to the affine parameter
$\lambda$. 
The components of the Riemann curvature with respect
to this frame diverge at the singularity, they are given by
\bea{cur-frame}
R_{2121} = -p^2 F, \quad\quad R_{3232} = p^2 F, \\ \nonumber
R_{2020} = -p^2 F, \quad\quad R_{3030} = p^2 F.
\eea

\subsection{Condition for a plane wave with a weak singularity}

When  a body falls into this singularity the tidal forces  diverge.
The tidal forces 
suffered by an object falling into the singularity in 
its parallel propagated  frame 
are given by the  curvature components in the time like directions in
\eq{cur-frame}. 
The singularity is weak if the rate of divergence is small enough so
that the object is not distorted as it hits the singularity. 
The distortion is proportional to the second integral of the tidal force
\cite{Ellis}. 
Consider the case of power law divergence at the origin $F(u) = 1/u^{\alpha}$,
the distortion is finite at the origin if $0<\alpha<2$

This notion was  made more precise by \cite{Tipler}. A singularity is defined to
be weak if for every time like geodesic which ends on it there exists  linearly
independent space like vorticity-free Jacobi fields along the geodesic normal to
its tangent vector which define a volume element that remains finite
at the singularity. This captures the intuitive notion of the distortion 
of the object being finite. We now verify that for the case of power law
divergence $F(u) = 1/u^{\alpha}$ with $0<\alpha<2$ the Jacobi fields define a
finite volume element at the singularity.

Jacobi fields satisfy the geodesic deviation equations, in the orthonormal frame
\eq{orth-frame} the space like Jacobi fields satisfy
\bea{geod-dev}
& &\frac{d^2 \eta^1}{d\lambda^2} = 0, \\ \nonumber
& &\frac{d^2 \eta^2}{d\lambda^2} - p^2 \frac{1}{u^\alpha} \eta^2 = 0, \\
\nonumber
& &\frac{d^2 \eta^3}{d\lambda^2} +p^2 \frac{1}{u^\alpha} \eta^3 = 0,
\eea
here $u= p\lambda$.
The geodesic deviation equation is diagonal, that is it does not mix the
space like Jacobi fields $\eta^i$ among each other, therefore the vorticity of
these Jacobi fields is zero. 
Solutions of  these equations 
for $0<\alpha<2$ are known in closed form.  Consider the time like
geodesic with $u =p\lambda$, 
an example of Jacobi
fields which remain finite at the origin is given by
\be{jaco-fin}
\eta^1 = \lambda + 1, \quad\quad
\eta^2 = \sqrt{p\lambda}K_{\nu} (2\nu (p\lambda)^\frac{1}{2\nu}), \quad\quad
\eta^3 = \sqrt{p\lambda}Y_{\nu} (2\nu (p\lambda)^\frac{1}{2\nu}),
\ee
where $\nu = 1/(2-\alpha)$ and $Y_\nu$ and $K_\nu$ are the Bessel function and
the modified Bessel function respectively.
From \eq{exp-ky}, \eq{exp-ki} and \eq{exp-yi}, it is
clear that these functions are finite at the origin.
Thus the volume defined by the product
$\eta^1\eta^2\eta^3$ is finite at the origin. 
Therefore for power law
divergences $F(u) = 1/u^\alpha$, the singularity is weak if $\alpha<2$.
Note that not only the volume defined by the Jacobi fields is finite for these
class of singularities but also each of the Jacobi field is finite at the
singularity.

\subsection{Scalars fields in a weakly singular plane wave}

We now show that among the class of weak singularities 
discussed in the previous section we can find a subclass for which 
the stress energy tensor of a massless scalar does not diverge as seen by 
any time like observer falling in to the singularity.
The evaluation of the stress energy tensor 
in the frame of the in falling time like observer  provides an invariant 
method of testing whether the scalar field will cause an  
infinite back reaction at the singularity or not. This method of testing
for back reaction of a scalar in the presence of 
null singularities is common in the literature, 
see for instance in \cite{Chandra}.

It is convenient to work in the Rosen coordinate system which are well behaved at
the singularity. The details of the various Rosen coordinate systems  
for the plane wave with $F(u) =1/u^\alpha$ and $0<\alpha<2$ 
are given in the appendix.
In this coordinates the metric for the plane wave is given by
\be{ro}
ds^2 = d\uu d\vv + f^2(\uu) d\x^2 + g^2(\uu) d\y^2,
\ee
where $f$ and $g$ are 
\be{def-fg}
f(\uu) = \sqrt{\uu} K_\nu(2\nu \uu^\frac{1}{2\nu} ), \quad\quad\quad
g(\uu) = \sqrt{\uu} Y_\nu(2\nu \uu^\frac{1}{2\nu} ).
\ee
These Rosen coordinates are well behaved at the singularity $u=0$, from the
expansions in \eq{exp-ky}, \eq{exp-ki} and \eq{exp-yi}, it is clear that they
are finite at the origin.
The massless scalar wave equation in the background \eq{ro} reduces to
\be{waveq}
\frac{1}{fg} \left( 
\partial_{\uu} ( 2fg\partial_{\vv} \phi) + \partial_{\vv}(2 fg
\partial_{\uu} \phi) + \partial_{\x}( \frac{g}{f}  \partial_{\x} \phi) + 
\partial_{\y}( \frac{f}{g} \partial_{\y} \phi)  \right) =0.
\ee
The above equations are invariant under translations in $\vv, \x, \y$, therefore
we can solve these equations by assuming the form 
$\phi = \phi(\uu) \exp( i\omega v + i k_x x + i k_y y)$. 
Substitution this form in \eq{waveq} we obtain
an ordinary equation for $\phi(u)$. The complete solution for $\phi$ is given by
\be{scalwave}
\phi(\uu, w, k_x, k_y) = \frac{1}{\sqrt{fg}} 
\exp\left(i\omega \vv + ik_x \x + ik_y \y + \frac{1}{4i\omega}
\int d\uu ( \frac{k_x^2}{f^2} + \frac{k_y^2}{g^2} ) \right).
\ee
Note that the wave function is well defined and smooth at the origin. The
nontrivial dependence of this wave function is in the $\uu$ coordinate. 
The derivative of the wave function with respect to $\uu$ is 
given by
\be{derscal}
\partial_\uu \phi = 
\left( -\frac{1}{2} \frac{\partial_\uu( fg)}{fg} + \frac{1}{4iw}(
\frac{k_x^2}{f^2} + \frac{k_y^2}{g^2}) \right) \phi. 
\ee
As $f$ and $g$ are well behaved at the origin
the behaviour of the derivative at the origin is dictated by the of logarithmic
derivative of the product $fg$ at the origin. From the expansions in
\eq{exp-ij}, \eq{exp-ky}, \eq{exp-ki} and \eq{exp-yi} it is clear that this
derivative is finite only for $0<\alpha<1$. The stress energy tensor is a
bilinear function of the derivatives of the scalar field, therefore we expect
the stress energy tensor evaluated in the frame of the in falling observer
will be well behaved for $0<\alpha<1$.

The general solutions for the scalar field is given by
\be{gen-sol}
\phi(\uu, \vv, \x, \y) = \int d w dk_x dk_y \phi(w, k_x, d_y) \chi(w, k_x, k_y)
\ee
Here $\chi(w, k_x, k_y)$ are the Fourier coefficients which are determined by
boundary conditions given on a constant $u$ surface. If the boundary conditions
are given in a patch which is beyond 
the validity of the Rosen coordinates \eq{ro}, then by solving
the wave equation in that patch corresponding to the boundary conditions 
and using the overlap
of patches one can determine $\chi(w, k_x, k_y)$ \footnote{One might have to go
through  several Rosen coordinate patches.}. 
For simplicity in our analysis 
we will assume our boundary conditions are such that $\chi(w', k_x, k_y) =
1/2(\delta(w-w') + \delta(w+w') )\delta(k_x)\delta(k_y)$. This is basically
assuming the scalar field is independent in $\x$ and $\y$ and it is a
cosine wave with frequency $w$. The analysis which we detail below can be easily
repeated for an arbitrary Fourier coefficient $\chi(w, k_x, k_y)$
The non-zero components of the stress energy tensor of the cosine wave 
$\phi(\uu,\vv) = \frac{1}{\sqrt{fg}} \cos(w\vv)$ are given by
\bea{stress-cos}
T_{\uu\uu} &=& \left( \frac{\partial_\uu(fg)}{fg} \right)^2 \cos^2(w\vv), \\
\nonumber
T_{\vv\vv} &=& w^2 \sin^2 (w\vv), \\ \nonumber
T_{\x\x} &=& \frac{f^2 w\partial_\uu(fg)}{2fg}  \cos(w\vv) \sin(w\vv), \\
\nonumber
T_{\y\y} &=& \frac{f^2 w \partial_\uu(fg)}{2fg}  \cos(w\vv) \sin(w\vv). 
\eea
All the components are well behaved at the origin if  $0<\alpha<1$. The stress
energy tensor for the sine wave is obtained by replacing the cosines by the
sines and vice-versa in the above formulae.

One might suspect that the reason that the stress energy tensor is well behaved
is due to the choice of nice coordinates near the singularity and a coordinate
artifact.  In fact if the
stress energy tensor given in \eq{stress-cos} is converted to the 
plane wave coordinates in  \eq{epwave} the
stress energy tensor is divergent. The natural way to check if the stress
energy tensor is divergent in a coordinate invariant manner is to
evaluate the components of
the stress energy tensor as seen in the frame of an observer traveling on 
any time like geodesic \cite{Chandra}.  We now show that that the stress
energy tensor of a scalar field seen by a in falling observer is finite in the
class of singular plane waves with $F(u) = 1/u^\alpha$ with $0<\alpha<1$.
The geodesic equations in Rosen coordinates are
given by
\bea{geod}
\frac{d^2 \uu}{d\lambda^2} &=& 0, \\ \nonumber
\frac{d^2 \x}{d\lambda^2} + 2\frac{\dot{f}}{f}\frac{d\uu}{d\lambda} 
\frac{d\x}{d\lambda} &=&0,
\\ \nonumber
\frac{d^2 \y}{d\lambda^2} + 2\frac{\dot{g}}{g}
\frac{d\uu}{d\lambda} \frac{d\y}{d\lambda} &=&0,
\\ \nonumber
\frac{d}{d\lambda} \left( 
\frac{d\vv}{d\lambda} + 
f^2 ( \frac{d\x}{d\lambda})^2 + g^2 (\frac{d\y}{d\lambda})^2
\right) &=&0.
\eea
The solutions for the tangent vector along the geodesic is given by
\bea{geodtgt}
& &\frac{d\uu}{d\lambda} = 1, \quad\quad 
\frac{d\x}{d\lambda} = \frac{a}{f^2}, \quad\quad  
\frac{d\y}{d\lambda} = \frac{b}{g^2}, \\ \nonumber
& &\frac{d\vv}{d\lambda} = -1- \frac{a^2}{f^2} - \frac{b^2}{g^2}
\eea
Here we have scaled $\lambda$ so that velocity in the $u$ direction is unity and 
normalized the tangent vector such that  its norm is $-1$. $a$ and $b$ are
arbitrary constants. Note that the velocity along the $\vv$ direction is finite
in the Rosen coordinates unlike the case in the plane wave coordinates as in
\eq{geo}.
Let the tangent vector along the geodesic be denoted by $v_0^\mu$. The other
vectors in the parallel propagated orthonormal tetrad along this geodesic are
\bea{ro-orth}
& &v_1 = 
(1, \;1- \frac{a^2}{f^2} -\frac{b^2}{g^2},\; \frac{a}{f^2}, \;\frac{b}{g^2} ),
\\ \nonumber
& &v_2 = ( 0, -\frac{2a}{f}, \frac{1}{f}, 0) , \\ \nonumber
& & v_3 = (0, -\frac{2b}{g} , 0, \frac{1}{g} ).
\eea
It is clear that since all the components of this tetrad are well behaved at
$\uu=0$ the stress energy tensor in this frame will be well behaved. 
For instance the $00$ component of the
stress energy tensor of the scalar field in the frame of the in falling
observer is given by
\bea{stess-in}
T_{00} &=& T_{\mu\nu} v^\mu_0 v^\nu_0, \\ \nonumber
&=& T_{\uu\uu} + T_{\vv\vv}\left(\frac{d\vv}{d\tau}\right)^2 + T_{\x\x} 
\left(\frac{d\x}{d\tau} \right)^2 +
T_{\y\y}\left(\frac{d\y}{d\tau} \right)^2.
\eea
It is clear from \eq{stress-cos} and  from \eq{geodtgt} the above expression for
$T_{00}$ is well behaved at $\uu=0$. Similarly from \eq{stress-cos} and 
\eq{ro-orth} the other components also will be well behaved at the singularity.
The apparent divergence of the stress energy tensor 
in the  plane wave coordinate given in \eq{epwave} 
is canceled when contracted with the velocity vectors of the
geodesic. Note that in \eq{geo}
the velocity vector along the $v$ direction diverges.

Thus for the subclass of plane wave singularities which diverge as $F(u)=
1/u^\alpha$ with $0<\alpha<1$ the stress energy tensor of a scalar field as seen
by an in falling geodesic does not diverge, therefore these backgrounds are
classically stable from large back reaction effects for a scalar probe.

\section{Strings in weak singular plane waves}

In this section we study string propagation in plane waves with weak
singularities. We 
show that that the classical 
string modes for plane waves with singularity of the form
$F(u) = 1/u^\alpha$ and $0<\alpha<1$  are well behaved at the singularity,
in fact even the derivative of the string modes are well behaved. This suggests a
natural method for 
extension of the metric in \eq{epwave} beyond the singularity $u=0$. 
Considering the metric as $F(u) = 1/|u|^\alpha$ 
One can match the classical string modes and its derivative at $u=0$ and
continue the modes smoothly across  $u=0$. 
We discuss this continuation of the string modes in  this section and 
will  provide more
justification for this extension of the metric in the  section 4.

Consider the metric given in \eq{epwave} embedded in bosonic string theory
\footnote{This discussion can be easily extended for the case of the
superstring.}.
The world sheet action of the bosonic string is given by
\bea{ws-act}
S &=& -\frac{1}{4\pi \alpha'} \int d \tau d\sigma \left(
\partial_a U \partial^a V + F(U) (X^2 -Y^2) \partial_a U \partial^a U 
\right. \\ \nonumber
&\quad&\quad\quad \quad + \left. 
\partial_a X \partial ^a X  + \partial _a Y \partial^a Y 
 +  \partial_a X^i \partial ^a X_i \right). 
\eea
Here $i =3, \ldots 25 $, we have used 
the Minkowski signature for the world sheet metric. The classical string modes
can be studied in the light cone gauge,  
substituting  $U = p \tau$ in the action \eq{ws-act} we 
obtain
\bea{ls-act}
S &=& -\frac{1}{4\pi \alpha'} \int d\tau d\sigma \left(
-p\frac{dV}{d\tau}  -p^2 F(p\tau) (X^2 -Y^2)  \right. \\ \nonumber
&\quad& \quad\quad\quad 
\left. + \partial_a X \partial ^a X  + \partial _a Y \partial^a Y 
+ \partial_a X^i \partial ^a X_i \right). 
\eea
The
coordinates $X^i$ are free and their solution is given by the usual mode
expansion, we will ignore these modes in our discussion. 
The non-trivial equations of motion are for the $X$ and $Y$ fields, they are
given by
\bea{non-eq}
\partial^a\partial_a X + p^2F(p\tau) X &=& 0,  \\ \nonumber
\partial^a \partial_a Y - p^2 F(p\tau) Y &=& 0.
\eea
Translations in the world sheet coordinate $\sigma$ is a symmetry of the above
equations, this and the fact that $\sigma $ is a periodic coordinate imply that
we can expand the modes as
\be{mod-exp}
X(\tau, \sigma) = \sum_{n=-\infty}^{n=\infty} X_n(\tau) 
e^{in\sigma} , \quad\quad\quad
Y(\tau, \sigma) = \sum_{n=-\infty}^{n=\infty} Y_n (\tau) e^{in\sigma} 
\ee
where $X_n$ and $Y_n$ satisfy the differential equations
\bea{xy-diff}
\ddot{X}_n + n^2 X_n - p^2F(p\tau) X_n &=& 0, \\ \nonumber
\ddot{Y}_n + n^2 Y_n + p^2 F(p\tau) Y_n &=& 0. 
\eea
Thus the equations for the classical modes reduce to equations for time
dependent oscillators with repulsive and attractive potentials.  
The classical modes for the coordinate $V$ is
determined by the following constraint equations
\bea{constr}
-p\dot{V} &=& (\dot{X})^2 + (\dot{Y})^2 + (X')^2 + (Y')^2  + F(p\tau) (X^2- Y^2),
\\ \nonumber
-p V' &=& 2 ( \dot{X} X' + \dot{Y} Y' ),
\eea
here the derivatives are with respect to the world sheet coordinates.

Let us now discuss the modes for  
singular plane waves with $F(u) =1/u^\alpha$ and 
$0<\alpha<1$. 
The zero modes for the $X$ and $Y$ coordinates 
are the geodesic equations for these coordinates in \eq{geo}, 
their solutions are given by
\bea{zero-mode}
X_0 &=& \alpha_0 \sqrt{p\tau} I_\nu \left( 2\nu (p\tau)^{1/2\nu} \right)
+ \tilde{\alpha}_0 \sqrt{p\tau} K_\nu \left( 2\nu (p\tau)^{1/2\nu} \right), \\
\nonumber
Y_0 &=& \beta_0 \sqrt{p\tau} J_\nu \left( 2\nu (p\tau)^{1/2\nu} \right)
+ \tilde{\beta}_0 \sqrt{p\tau} Y_\nu \left( 2\nu (p\tau)^{1/2\nu} \right),
\eea
where $\nu = 1/(2-\alpha)$ and $\alpha_0, \tilde{\alpha}_0, \beta_0,
\tilde{\beta}_0$ are arbitrary constants. $J_\nu$, $Y_\nu$ and $I_\nu$, $K_\nu$
are the Bessel functions and modified Bessel functions respectively.
Note that for $0<\alpha<1$, $\nu$ 
always lies between $0$ and $1$, and therefore $\nu$ never takes on integral
values. From the expansion of these 
Bessel functions given in 
\eq{exp-ij} and \eq{exp-ky} it is easy to see that they are finite and their
derivatives are also finite at the origin for $0<\alpha<1$.

Consider the non zero modes for the coordinate $X_n$, the behaviour of these
modes can at $u=0$ can be studied by substituting $X_n = x_n h$
in \eq{xy-diff} with 
$h(p\tau) = \sqrt{p \tau} I_{-\nu} \left( 2\nu ( p\tau)^{1/2\nu} \right) $ the
equations of motion for $X_n$  then reduces to
\be{xn-eq}
\ddot{x}_n + 
2\dot{x}_n \frac{1}{h(p\tau)} \frac{d}{d\tau} h(p\tau)  + n^2 x_n = 0,
\ee
where we have used the fact that $h$ solves the equation
\be{eqoh}
\ddot{h} + \frac{1}{\tau^\alpha} h =0.
\ee
The expansion for $h$ is given by
\be{exp-h}
h(\tau) = \nu^{-\nu} \left( \frac{1}{\Gamma( -\nu+1)} + \frac{\nu^2
\tau^{2-\alpha}}{1!\Gamma(-\nu+2)} + \frac{\nu^4 \tau^{2(2-\alpha)}}{2!\Gamma(
-\nu+3) } + \cdots \right),
\ee
it is clear from the above expansion that the logarithmic derivative of $h$ goes
to zero at $\tau=0$ if $0<\alpha<1$. Therefore from \eq{xn-eq} the two linearly
independent solutions for $X_n$ as $\tau \rightarrow 0$ are given by
\be{sol-xn}
X_n = e^{\pm in\tau}\sqrt{p\tau}
I_{-\nu} ( 2\nu (p\tau)^{\frac{1}{2\nu}}) 
\left( 1 + O(\tau^{2-\alpha}) \right),
\quad\quad \hbox{as} \quad\quad \tau \rightarrow 0.
\ee
The fact that these solutions are linearly independent can be shown by the
evaluation of the Wronskian. 
Thus the modes are well behaved at $\tau=0$ furthermore 
from \eq{sol-xn} it is also clear that the derivatives are also well behaved at
the origin.
A similar argument for the modes $Y_n$ shows that they are given by
\be{sol-yn}
Y_n = e^{\pm in\tau}\sqrt{p\tau} 
J_{-\nu} ( 2\nu (p\tau)^{\frac{1}{2\nu}} )
\left( 1 + O(\tau^{2-\alpha}) \right) , 
\quad\quad \hbox{as} \quad\quad \tau \rightarrow 0.
\ee
Thus the modes for the $Y_n$ and its derivative are also well behaved at
the origin.
We have also checked the 
the fact that these modes are well behaved at the origin 
by a Frobenius expansion which can be performed for 
rational values of $\alpha$ and numerically. Note that the in determining the
behaviour of the modes the crucial role was played by the change of variable
from $X_n$ to $x_n$. This transformation is precisely the transformation to a
specific Rosen coordinates as can be seen from \eq{cortrans}.
The modes of the coordinate $V$ is determined from the constraints 
\eq{constr}, it is easy to see from these equations that $V$ is finite but
$\dot{V}$ in general diverges at the origin.

We now outline a  natural completion of the plane wave metric with 
a power law divergence in $F(u) =
1/u^\alpha$ and $0<\alpha<1$ to the region $u<0$. Consider the extension to be
the plane wave metric with $F(u) = 1/|u|^\alpha$ at the origin, 
then if the two linearly independent solutions for $X_n$ are 
$X_n^{(1)}(\tau)$ and $X_n^{(2)}(\tau)$ for $\tau>0$, the two linear independent
solutions for $\tau<0$ are given by $X_n^{(1)}(-\tau)$ and $X_n^{(2)}(-\tau)$.
We have seen that for $0<\alpha<1$ these modes and their derivatives are well
behaved at the origin, therefore one can match the value of the modes and their
derivative for the solutions in the two regions at the origin and naturally
continue them across the singularity. This procedure cannot
be done for the case of singularities with $1\le\alpha<2$ as the derivative of
the modes diverge at the origin. 
The modes for the world sheet coordinate $V$ is determined from the 
constraints given in \eq{constr}. 
Note that in this completion of the metric the zero modes as well as the
non-zero modes of the string are in equal footing, unlike the case of the usual 
orbifolds in which geodesics or the zero modes cannot be completed while the
non-zero modes of the string can be extended \footnote{The author thanks Ashoke
Sen for pointing this out.}.

\section{Tests for the extended metric}

We have seen in the previous section 
that for plane waves with singularity $F(u) = 1/u^\alpha$ that there is a  
natural
extension of the metric to $u<0$ with $F(u) = 1/|u|^\alpha$. One can ensure that
all the string modes are continuous and differentiable at the origin for
$0<\alpha<1$. This still does not guarantee that the metric is singularity free
from the point of view of string theory. Though plane wave metrics do not
exhibit particle creation there is a phenomenon of mode creation, a string
passing through the singularity can become excited, for the metric to be
non-singular from the point of view of string theory it is important to show
that mode creation is finite. For instance with $F(u) = \delta(u)$
though the modes can be continued across the singularity (the derivative is
discontinuous) the string gets infinitely excited as it passes through this
singularity. This infinte excitation is infact due to the infinite
extent of the shock wave profile  and it similar to the case of
$\alpha =1$ \cite{deVega:1992ke,deVega:1990nr} 

In this section we show that the mode creation in the continued metric with
$F(u) = 1/|u|^{\alpha}$ and  $0<\alpha<1$ is finite, this was noticed in
\cite{deVega:1992ke}. 
The extension of the metric to $u<0$ will also be natural if there exists a
family of smooth metrics which are at a finite distance 
from the metric $F(u) = 1/|u|^\alpha$ in the space of
metrics. Then this extended metric is just a point in the moduli space, much
like the case of the $R^4/Z_2$ orbifold, the singular $R^4/Z_2$ metric is
just a limit of the smooth Eguchi-Hansen space.
We construct such a family of smooth metrics for the extended metric.
We also show that there is a well defined
operator in the string sigma model which resolves the singularity.
Thus this class of plane wave metrics can be smoothened out much like the 
$R^4/Z_2$ orbifold singularity.

\subsection{Mode creation}

The world sheet theory for the string in these time dependent backgrounds has
time dependent potentials. Time dependent potentials in general will mix
positive frequencies and negative frequencies  on the world sheet, which implies 
there is a transition amplitude between the oscillator modes of the vacuum at
$\tau = -\infty$ to the oscillator modes of the vacuum at $\tau =
-\infty$, here $\tau$ stands for the world sheet time. 
The mixing of modes will in general excite  string modes as it passes through the
time dependent background.
To compute the amplitude for mode creation through a time dependent back ground
one needs to know the solution for the differential equations \eq{xy-diff}.
Solutions for these equations do not exist in closed form \footnote{ For the
case $\alpha=1$ solutions exist in closed form.}, 
we therefore resort to approximate methods. 
Since we are interested in the behaviour close to
the singularity our methods should be valid in that region. 
It is easy to see that for $\alpha <2$ the WKB method is not valid
close to  $u=0$, the other approximate method is perturbation theory.
In \cite{deVega:1992ke} the (mass)$^2$ of the excited modes for the 
case strings passing through plane waves 
singularities with $0<\alpha<1$ was estimated using
first order perturbation theory and show to be finite.
Here we emphasize the validity doing perturbation theory 
in spite of the presence of the singularity at $u=0$ 
We estimate the mode creation to the second order in perturbation theory and
show that string does not get infinitely excited as it passes through the
singularity. 

As we are only interested in the singularity at the origin we can assume that
the function $F(u) = \frac{1}{|u|^\alpha}$ is modulated by a function which
is smooth and falls of at infinity. 
To keep the calculations involved simple  
we perform the analysis  below with $F(u)= 1/|u|^\alpha$
for $|u|<T$ and $F(u) =0$ for $|T|>0$ sufficiently large but finite.
We set up the perturbation theory for the modes of the coordinate $X$.
A similar treatment will go through for the coordinate $Y$.
The solution of equations of motions for the world sheet coordinate 
$X$ is given in terms of the following series
\bea{sol-x-ser}
X_n (\tau)&=& X_n^{(0)}(\tau)  + \int d\tau_1 G(\tau -\tau_1)
p^2F(p\tau_1)X_n^{(0)}(\tau_1) \\
\nonumber
&+& \int d\tau_2 d\tau_1 G(\tau-\tau_2) p^2F(p\tau_2) G(\tau_2 -\tau_1) p^2
F(p\tau_1) X_n^{(0)}(\tau_1) +  \ldots
\eea
All the integrals run from $-\infty$ to $\infty$.
The expansion parameter can be thought of as the light cone momentum $p$. 
$X_n^{(0)}$ is the solution of the unperturbed equation 
\be{unperteq}
\ddot{X_n}^{(0)} + n^2 X_n^{(0)} = 0, 
\ee
and  $G(\tau-\tau')$ is its Green's function which is given by
\be{greenbc}
G(\tau-\tau') = \frac{\theta(\tau'-\tau)}{2in} \left( e^{in(\tau-\tau')} -
e^{-in(\tau -\tau')} \right),
\ee
where $\theta(\tau-\tau')$ is the step function which ensures that the above
Green's function vanishes for $\tau>\tau'$. We require this boundary condition
as we are propagating the solutions from $\tau =\infty$ to 
$\tau= -\infty$.
It is clear from the perturbative expansion in \eq{sol-x-ser} all the integrals 
are well defined if  $F(u) = 1/|u|^\alpha$ with $0<\alpha<1$. 
This is because the interaction potential $F(p\tau)$ always occurs with an
integration in $\tau$, so the integrals around $u=0$ are well defined
\footnote{For $F(u) = \delta(u)$ the perturbation expansion truncates at the
first order giving the exact answer.}.
Note that it is not clear that one can trust the perturbative analysis for
$\alpha \ge 1$.

For mode creation what is important is to obtain the amplitude of the incoming
wave after scattering from the potential. 
Consider the incoming wave at $\tau =\infty$ as $e^{i n \tau}$,
then the amplitude of the coefficient of the outgoing wave $e^{-in\tau}$ at 
$\tau=-\infty$
is given by
\be{in-sc}
B_ne^{-in\tau} = \lim_{\tau \rightarrow -\infty} \int d\tau' \left( \tilde G(\tau
-\tau') - \frac{1}{in} \frac{d}{d\tau} \tilde G(\tau -\tau') \right) V(\tau')
e^{in\tau'}. 
\ee
Here $B_{n}$ refers to the coefficient of the outgoing wave and
$\tilde{G}$ stands for full Green's function whose expansion is given by
\be{green-full}
\tilde{G} (\tau -\tau') = G(\tau - \tau') + \int d\tau_1 G(\tau
-\tau_1)p^2F(p\tau_1) G(\tau -\tau') + \ldots
\ee
Evaluating $B_n$ to the  first order in perturbation theory gives
\be{first-or}
B^{(1)}_n = \frac{\hat{F}( -2n) }{2in}
\ee
Here $\hat{F}$ is the Fourier transform of the potential defined by
$ \hat{F}(w) = \int dt p^2 F(pt) e^{iwt}$.
For the potential $F(u) = 1/u^\alpha$ with $0<\alpha<1$ it is given by
\be{ft-pot}
\hat{F}(w) =  2 \frac{p^{2-\alpha}}
{|w|^{1-\alpha} }\Gamma(1-\alpha) \sin(\frac{\pi\alpha}{2}) + 
O(\frac{1}{wT^\alpha}),
\ee
where the $O(1/wT^\alpha)$ terms refers to the corrections due to the 
fact that the potential is non zero only for 
$ |u|<T$.
From the Bogoliubov transformations for the oscillator modes 
the expectation value of the number operator at $\tau = -\infty$ of the vacuum
at $\tau= \infty$ is given by
\be{btrans}
\langle 0_+| N_n | 0_+ \rangle = |B_n|^2,
\ee
where $N_n$ refers to the number operator  of either the left movers or the
right movers in the $n$th level. 
The expectation value of the number operator can be translated to 
the expectation value of the mass of
the excited string after it passes through the singularity 
which is given by 
\be{mass-ex}
<M^2> = \frac{4}{\alpha'}( 2\sum_n   n <N_n^i>  -2).
\ee
From the equations  \eq{first-or}, \eq{ft-pot} and \eq{btrans} we see the mass
contribution from the excited modes corresponding to the $X$ coordinate 
in first order perturbation theory is given by 
\be{tot-mass}
<M^2_x> \sim \frac{2}{\alpha'} \sum_n \frac{1}{n^{3-2\alpha}}.
\ee
It is easy to see that this sum converges for $0<\alpha<1$. A similar analysis
holds for the $Y$ coordinate, thus the (mass)$^2$ of the excited string remains
finite in first order perturbation theory. 
To show that the leading estimate for the amplitude for mode creation is given
by the first order term we evaluate the 
Bogoliubov coefficient $B_n$ to the next order in perturbation theory.
This is given by
\bea{sec-or2}
B^{(2)}_n &=& 
\frac{1}{2n^2}  \left( \hat{F}(-2n) \hat{F} (0)   -
\hat{F}(-4n) \hat{F}(-2n) \right), \\ \nonumber
&=& \frac{p^{2(2-\alpha)}}{2n^2(n^{1-\alpha}) } \left[
\frac{2T^{1-\alpha}}{1-\alpha} - \left( \frac{1}{4n} \right)^{1-\alpha} \right]
+ O(\frac{1}{nT^\alpha}), 
\eea
note that $B_n^{(2)}$  
is suppressed by higher powers of $n$ and is finite for a
given $T$, therefore the amplitude for mode creation given in \eq{tot-mass} is
the leading estimate.

\subsection{Nonsingular metrics close to the extended  metric} 

In this section we show that there exists a one parameter 
family of smooth metrics which are at a finite distance from 
plane wave metrics given in \eq{epwave} with a
singularity of the form $F(u) = 1/|u|^\alpha$ with $0<\alpha<1$ 
at the origin.
The natural way to find a smooth metric which reduces to this singular metric in
a limit is to deform the function $F(u)$ to any smooth function at $u=0$. For
instance the function $F(u) = (u^2 + a^2)^{-\alpha/2}$ is a possible candidate.
The difficulty in considering these kind deformations is that 
they are all null deformations, that is their distance in the
space of metrics is null. 
This is easily seen as follows, 
the distance in the space of metrics is given by
\be{dist-met}
\Delta = \int d^{4}x \sqrt{g} g^{\mu\nu} g^{\rho\sigma} \delta g_{\mu\rho} \delta
g_{\nu\sigma},
\ee
where $\delta g_{\mu\nu}$ refers to the change in the metric. Note that the
metric we have in \eq{epwave} is such that $g^{\mu\nu}$ is zero if $\mu$ or
$\nu$ refers to the $u$ direction. Therefore deformations  involving just the
change in the function $F(u)$ is null. To obtain deformations which smooth out
the singularity one should involve the transverse directions $x$ and $y$, this
is conveniently done in the
Rosen coordinates,  in these coordinates 
the metric in \eq{epwave} is given by
\be{ro-ex}
ds^2 = d\uu d\vv + f(\uu) d\x^2  + g( \uu) d\y^2.
\ee
The details of the transformation from the usual plane wave coordinates to the
Rosen coordinates are given in \eq{cortrans} and \eq{difftrans}. If 
the singularity in $F(u)$ is of the form $1/|u|^\alpha$ with $0<\alpha<1$
then for the patch containing $\uu=0$, 
the functions $f$ and $g$ can be chosen to be
\be{fg-full}
f(\uu) = \sqrt{|\uu|} K_\nu( 2\nu|\uu|^{\frac{1}{2\nu} }), \quad\quad\quad
g(\uu) = \sqrt{|\uu|} Y_\nu( 2\nu|\uu|^{\frac{1}{2\nu} }). 
\ee
We can construct 
a one parameter family of metrics which are not singular at $u=0$ 
as follows. Consider the metric 
\be{family}
ds^2 = d\uu d\vv + f^2( \uu +a ) d\x^2 + g^2(\uu +a) dy^2,
\ee
where $a\ge 0$.
These are solutions of the equations of motion if 
 $f$ and $g$ satisfy the following equations
\be{cond-sol}
\frac{d^2}{d\uu^2} f(\uu +a) = F(\uu +a) f(\uu +a), \quad\quad
\frac{d^2}{d\uu^2} g(\uu +a) = -F(\uu +a) g(\uu +a), \quad\quad
\ee
here $F(\uu +a) = (|u| + a)^{-\alpha}$.
Therefore in the patch containing $u=0$,  $f$ and $g$ are given by
\be{fg-shift}
f(\uu+a) = \sqrt{|\uu|+a} K_\nu( 2\nu(|\uu|+a )^{\frac{1}{2\nu} }), \quad\quad
g(\uu+a) = \sqrt{|\uu|+a } Y_\nu( 2\nu(|\uu|+a)^{\frac{1}{2\nu} }). 
\ee
When $a=0$
this family of metrics reduces to the singular metric characterized by 
$F(u)= 1/|u|^\alpha$.
The metrics in \eq{family} are not singular, the curvature components are
proportional to $F(u + a)$ which does not diverge at the origin.

We now show that the one parameter deformation constructed above is close by to
the singular extended metric by evaluating the Zamolodchikov metric of the
operator responsible for the deformation in the string sigma model. 
The string sigma model for the metric in \eq{ro-ex} is given by
\be{ro-sigm}
S= -\frac{1}{4\pi\alpha'} \int d\tau d\sigma \left(
\partial_a U \partial^a U  + f^2 \partial_a X \partial^a X + g^2 \partial_a Y
\partial^a Y + \partial_a X^i \partial^a X_i \right).
\ee
For very small $a$, the deformation of the metric can be obtained by a Taylor
expansion of the function $f(\uu +a)$ and $g(\uu +a)$ and retaining the first
order term. Then operator which is responsible for this 
infinitesimal  deformation in the string
sigma model is given by
\be{oper}
 O(\tau, \sigma) = 
a\left( 2f \frac{d f}{d\uu} \partial_a  X \partial^a X 
+ 2g \frac{dg} {d\uu} \partial_a Y \partial^a Y \right). 
\ee
The Zamolodchikov metric $G_{OO}$ or the norm  of this operator is defined by
\be{z-met}
G_{OO}\, a^2 = \lim_{\epsilon\rightarrow 0} \epsilon^4 \langle
{\cal O} ( \tau, \sigma+ \epsilon)
{\cal O} ( \tau, \sigma) \rangle.
\ee
Note that from the expansions of the functions $f$ and $g$ in the appendix this
operator is well defined for $0<\alpha<1$.
To evaluate the Zamolodchikov metric it is 
convenient  to appeal to the formula relating it to the
Weil-Petersson metric in the space of metrics 
\cite{Candelas:1990qn} which is given by
\be{zwp-rel}
G_{OO}\, a^2 
= \frac{1}{V} \int d^{26} x \sqrt{g} g^{\mu\nu}
g^{\rho\sigma} \delta g_{\mu\rho} \delta g_{\nu\sigma},
\ee
here $g_{\mu\nu}$ stands for the undeformed metric with $a=0$ given 
in  \eq{ro-ex}, $\delta g_{\mu\nu}$ stands for the infinitesimal deformation
in the metric by the operator \eq{oper} and $V = \int \sqrt{g}d^{26} x$.
After substituting for these from \eq{ro-ex} and \eq{oper} and canceling out
the volume of the remaining spectator dimensions we obtain
\be{valzmet}
 G_{ OO}
= \frac{4}{\int d\uu \sqrt{ f^2 g^2}} \int d \uu
 \sqrt{f^2 g^2} \left[
\left( \frac{1}{f}\frac{d f}{du} \right)^2 + 
\left(\frac{1}{g} \frac{d g}{du} \right)^2 \right].
\ee
As we are only interested in the singularity near the origin  the metric in
\eq{ro-ex} can be taken to be that of flat space for $|u|>T$. The
integrand in the numerator of above equation is supported over a compact
region in $u$. We render the denominator finite by integrating over a large but
finite region in $u$. \footnote{Note that in \cite{Candelas:1990qn} the
Zamolodchikov metric or the Weil-Pettersson metric was evaluated for 
deformations in Calabi-Yau
manifolds which is compact, therefore there was no need for this regularization.}
The only possible divergence can be at $u=0$, in that region the functions
reduce to those given in \eq{fg-full}. The integrand in \eq{valzmet} contains the
logarithmic derivatives of these functions  and 
from the expansions of these function at $u=0$ it is easy to see 
$G_{OO}$ is finite for $0<\alpha<1$. Therefore the infinitesimal deformation in
\eq{oper} can smooth out the singularity at $u=0$.

\section{Conclusions}

We have seen that plane wave solutions of pure gravity with weak singularities
i.e.  $F(u)= 1/u^\alpha$ and $0<\alpha<2$,
admit the  sub-class  with $0<\alpha<1$,
which do not exhibit large back reaction in the presence of
scalar probes. 
For these class of plane waves
classical string modes do not diverge
at the singularity. There is a natural way to continue the metric across the
singularity, this extended metric admits string propagation without the string
becoming infinitely excited. 
This extended metric lies at a finite distacne  to a family of smooth metrics in
the space of metrics. The operator which smoothes out the singularity has a
finite norm.
It will be interesting to see if in general there are classes of weak
singularities in which energies of scalar probes do not diverge and 
which admit smooth string propagation. The singularities studied in this paper
were null singularities, 
and it is of interest to see if there are similar weak time like
singularities. 
In plane waves with singularities which are 
supported by  background Neveu-Schwarz $B$-fields 
mode creation can be suppressed due to the coupling of the
$B$-field on the world sheet \cite{Maeda:2000pp}, here the interplay of the 
singular effect of  the geometry is effectively canceled by 
and that of the $B$-field. This is 
phenomenon is potentially of interest to understand  mechanisms
for resolutions of time dependent singularities.

The fact that weak singularities admit space like
Jacobi fields which define a finite volume element implies that  they can be
used to define a well behaved 
synchronous coordinate system near the singularity.
For the plane wave analyzed in this paper these were the Rosen coordinates.
One can then
impose the requirement for smooth behaviour of scalar and string probes in this
synchronous
coordinate system and translate this  to a condition on the Jacobi
fields. This will provide a useful general condition for 
classification of metrics which admit a weak singularity. 

\noindent
{\emph{Note added:}} After this work was completed the author noticed
\cite{Sanchez:2003ek} which has a partial overlap with this paper.

\acknowledgments

It is a pleasure to thank Mathias Blau, Martin O' Loughlin, K. S.  Narain, 
Tapobrata Sarkar, 
Ashoke Sen and Marija Zamaklar for useful  discussions. The author thanks the
organizers of PASCOS '03 for the opportunity to present this work at the
conference and the high energy theory group at the 
Tata Institute for Fundamental Research, Mumbai  
for hospitality during which part of the work was carried out. The work of the
author is supported in part by EEC contract EC HPRN-CT-2000-00148.

\appendix

\section{Rosen coordinates}

The plane wave metric given in \eq{epwave} is written in 
Brinkman coordinates which are global. 
Near the singularity it is convenient to use Rosen coordinates. 
In fact there is a choice of coordinates such that the metric is well defined at
the singularity if it is weak. To convert to Rosen coordinates we perform the
the following change of coordinates $(u, v, x, y) \rightarrow (\uu, \vv,
\x, \y)$ 
\bea{cortrans}
u &=& \uu, \\ \nonumber
v &=& \vv - \frac{df}{d\uu} f \x^2 - \frac{dg}{d\uu} g \y^2, \\ \nonumber
x &=& f \x, \\ \nonumber
y &=& g \y,
\eea
where $f$ and $g$ satisfy the equations
\be{difftrans}
\frac{d^2f}{d\uu^2} = F(\uu) f, \quad\quad 
\frac{d^2g}{d\uu^2} = -F(\uu) g. 
\ee
Under this coordinate transformation the plane wave metric in \eq{epwave} reduces
to 
\be{rosen}
ds^2 = d\uu d\vv + f^2 d\x^2 + g^2 d\y^2.
\ee
The general solutions for $f$ and $g$ in  \eq{difftrans}
for weak singularities with $F(u) = 1/u^\alpha$ and $0<\alpha<2$ is given by
\bea{deffg}
f(\uu) &=& A \sqrt{\uu} 
I_{\nu } (2\nu \uu^{\frac{1}{2\nu} } )
+ B \sqrt{\uu} 
K_{\nu} (2\nu \uu^{\frac{1}{2\nu} } ), \\ \nonumber
g (\uu) &=& C \sqrt{\uu} 
J_{\nu } (2\nu \uu^{\frac{1}{2\nu} } )
+ D \sqrt{\uu} 
Y_{\nu} (2\nu \uu^{\frac{1}{2\nu} } ), 
\eea
where $\nu = 1/(2-\alpha)$ and the functions $J_\nu, Y_\nu$ and $I_\nu, K_\nu$
refer to the Bessel functions and the modified Bessel functions  as
defined in \cite{Stegun}.  One can obtain different Rosen coordinate systems
corresponding to the various choices of the constants $A, B, C$ and $D$. 
It is useful to write out the expansions of the functions appearing in
\eq{deffg} around the origin, for the functions involving $I_\nu$ and $J_\nu$ 
the expansion are given by
\bea{exp-ij}
\sqrt{\uu} I_\nu( 2\nu \uu^\frac{1}{2\nu}  ) = 
\nu^\nu \uu \left( \frac{1}{\Gamma( \nu +1)}
+  \frac{\nu^2\uu^{2-\alpha}}{1! \Gamma ( \nu +2)}
+  \frac{ \nu^4 \uu^{2(2-\alpha)} }{2!\Gamma ( \nu + 3) } + \cdots \right), 
\\ \nonumber
\sqrt{\uu} J_\nu( 2\nu \uu^\frac{1}{2\nu}  ) = 
\nu^\nu \uu \left( \frac{1}{\Gamma( \nu +1)}
-  \frac{\nu^2\uu^{2-\alpha}}{1!\Gamma ( \nu +2)}
+ \frac{\nu^4\uu^{2(2-\alpha)} }{2!\Gamma ( \nu + 3) } - \cdots \right) . 
\eea
For the functions involving $K_\nu$ and $Y_\nu$ and when $\nu$ is not an integer
the expansions are given by
\bea{exp-ky}
\nonumber
\sqrt{\uu} K_\nu( 2\nu \uu^\frac{1}{2\nu} ) &=& 
\frac{\pi}{2\sin(\nu\pi)} \nu^{-\nu} \left( \frac{1}{\Gamma( -\nu +1)}
+  \frac{\nu^2\uu^{2-\alpha}}{1! \Gamma ( -\nu +2)}
+ \frac{ \nu^4\uu^{2(2-\alpha} }{2!\Gamma ( - \nu + 3) } + \cdots \right)  \\
\nonumber
&- & \frac{\pi}{2\sin(\nu\pi)}
\sqrt{\uu} I_\nu( 2\nu \uu^\frac{1}{2\nu} ),  \\ 
\sqrt{\uu} Y_\nu( 2\nu \uu^\frac{1}{2\nu} ) &=& 
\frac{-1}{\sin(\nu\pi)} \nu^{-\nu} \left( \frac{1}{\Gamma( -\nu +1)}
-  \frac{\nu^2\uu^{2-\alpha}}{1! \Gamma ( -\nu +2)}
+  \frac{ \nu^4\uu^{2(2-\alpha)} }{2!\Gamma ( - \nu + 3) } - \cdots \right),  
\nonumber \\
&+ & \cot{(\nu\pi)}
\sqrt{\uu} J_\nu( 2\nu \uu^\frac{1}{2\nu} ).
\eea
Note that the above functions are finite at the origin.
When $1\le \alpha<2$ there is a possibility  of
$\nu$ being an integer, then there
are logarithmic terms in the expansion for $K_\nu$ and $J_\nu$, still these
functions are finite at the origin,  
the logarithmic terms and the first few terms in the expansion 
of $K_n$ is given by
\bea{exp-ki}
\sqrt{\uu} K_n( 2n \uu^\frac{1}{2n} ) &=& (-1)^{n+1}  \sqrt{u}I_n( 2n
\uu^{\frac{1}{2n} } ) \ln( n \uu^\frac{1}{2n} ) \\ \nonumber
&+& \frac{n^{-n}}{2}\left( (n-1)! - \frac{(n-2)! n^2\uu^{(2-\alpha)} }{2!} +
\cdots (-1)^{n-1}
\frac{n^{2(n-2)} \uu^{(n-1)(2-\alpha)} }{(n-1)!}\right) \\ \nonumber
&+& (-1)^n \frac{n^n}{2} \uu \left( 
\frac{( \psi(1) + \psi(n+1) )}{n!}
+ \frac{( \psi(2) + \psi(n+2) )n^2\uu^{2-\alpha} }{2!(n+2)!} + \cdots
\right),
\eea
here $n = 1/(2-\alpha)$ is an integer and $\psi$ is the digamma function defined
\cite{Stegun}. Similarly the expansion for $Y_n$ is given by
\bea{exp-yi}
\sqrt{\uu} Y_n( 2n \uu^\frac{1}{2n} ) &=& \frac{2}{\pi} \sqrt{u}J_n( 2n
\uu^{\frac{1}{2n} } ) \ln( n \uu^\frac{1}{2n} ) \\ \nonumber
&-& \frac{n^{-n}}{\pi}\left( (n-1)! + \frac{(n-2)! n^2\uu^{(2-\alpha)} }{2!} +
\cdots \frac{n^{2(n-2)} \uu^{(n-1)(2-\alpha)} }{(n-1)!}\right) \\ \nonumber
&-& \frac{n^n}{\pi} \uu \left( 
\frac{( \psi(1) + \psi(n+1) )}{n!}
- \frac{( \psi(2) + \psi(n+2) )n^2\uu^{2-\alpha} }{2!(n+2)!} + \cdots
\right).
\eea
From the above expressions in \eq{exp-ky}, \eq{exp-ki} and \eq{exp-yi} it is
clear that for $0<\alpha<2$ the functions in \eq{deffg} involving $K_\nu$ and
$Y_\nu$ are finite at the origin and from \eq{exp-ij} the functions in
\eq{deffg} involving $I_\nu$ and $J_\nu$  vanish at the origin.
The Jacobian
for transformation to the Rosen coordinates is given by the product $fg$,
therefore to have a well defined coordinate system in the patch which contains
$u=0$,  $B$ and $D$ is not zero. 
It is convenient to choose $A=B=0$ in the Rosen coordinates in this patch.

\bibliographystyle{utphys}
\bibliography{sing}
\end{document}